\documentclass[useAMS,usenatbib,usegraphicx,referee]{mn2e}

\usepackage{amssymb}

\title[Wide-angle outflow and jet in HW2]{A wide-angle outflow with the simultaneous presence of
a high-velocity jet in the high-mass Cepheus A HW2 system}
\author[Torrelles et al.]{J. M. Torrelles$^{1}$\thanks{E-mail: torrelles@ieec.cat (JMT); npatel@cfa.harvard.edu (NAP); scuriel@astroscu.unam.mx (SC); robert@am.ub.es (RE);  jfg@iaa.es (JFG); l.rodriguez@crya.unam.mx (LFR); guillem@iaa.es (GA); wouter@astro.uni-bonn.de (WV); guido@das.uchile.cl (GG); raga@nucleares.unam.mx (ACR); pho@asiaa.sinica.edu.tw (PTPH)},
N. A. Patel$^{2}$, S. Curiel$^{3}$, R. Estalella$^{4}$, J. F. G\'omez$^{5}$, 
\newauthor
L. F. Rodr\'{\i}guez$^{6}$, J. Cant\'o$^{3}$, 
G. Anglada$^{5}$, W. Vlemmings$^{7}$, G. Garay$^{8}$,
\newauthor
A. C. Raga$^{9}$, and P. T. P. Ho$^{10,2}$
\\
$^{1}$Instituto de Ciencias del Espacio (CSIC)-UB/IEEC, Universitat de Barcelona, Mart\'{\i} i Franqu\`{e}s 1, E-08028 Barcelona, Spain\\
$^{2}$Harvard-Smithsonian Center for Astrophysics, 60 Garden Street, Cambridge, MA 02138, USA\\
$^{3}$Instituto de Astronom\'{\i}a (UNAM), Apartado 70-264, 04510 M\'exico
D. F., M\'exico\\
$^{4}$Departament d'Astronomia i Meteorologia and Institut de Ci\`{e}ncies del Cosmos (IEEC-UB), Universitat de Barcelona,\\~~Mart\'{\i} i Franqu\`{e}s 1, E-08028 Barcelona, Spain\\
$^{5}$Instituto de Astrof\'{\i}sica de Andaluc\'{\i}a (CSIC), Apartado 3004, E-18080 Granada, Spain\\
$^{6}$Centro de Radioastronom\'{\i}a y Astrof\'{\i}sica (UNAM), Morelia 58089, M\'exico\\
$^{7}$Argelander-Institut f\"ur  Astronomie, University of Bonn, Auf dem H\"ugel 71, D-53121 Bonn, Germany \\
$^{8}$Departamento de Astronom\'{\i}a, Universidad de Chile, Casilla 36-D, Santiago, Chile\\
$^{9}$ Instituto de Ciencias Nucleares (UNAM), Apartado 70-543, 04510 M\'exico D. F., M\'exico  \\
$^{10}$Academia Sinica Institute of Astronomy and Astrophysics, Taipei, Taiwan}
\begin{document}

\date{MNRAS, in press}

\pagerange{\pageref{firstpage}--\pageref{lastpage}} \pubyear{2010}

\maketitle

\label{firstpage}

\begin{abstract}
We present five epochs of VLBI water maser observations around the
massive protostar Cepheus A HW2 with  0.4 mas (0.3 AU) resolution. The main
goal of these observations was to follow the evolution of the
remarkable water maser linear/arcuate structures found in earlier
VLBI observations. Comparing  the data of our new epochs of  observation with
those observed five years before, we find that at ``large'' scales of
$\ga$1$''$ (700 AU) the  main regions of maser emission persist, implying that both the
surrounding medium and the exciting sources of the masers have been
relatively stable during that time span. However, at smaller scales
of $\la$ 0.1$''$ (70 AU) we see large changes in the maser structures,
particularly in  the expanding arcuate structures R4 and R5. R4
traces a
nearly elliptical patchy ring of $\sim$ 70 mas size (50 AU)
with expanding motions of $\sim$ 5~mas~yr$^{-1}$ (15~km~s$^{-1}$),
consistent with previous results of Gallimore and collaborators.
This structure is probably driven by the wind of a still unidentified
YSO located at the centre of the ring ($\sim$ 0.18$''$ south of
HW2). On the other hand, the R5 expanding bubble structure (driven
by the wind of a previously identified YSO located $\sim$ 0.6$''$
south of HW2) is currently dissipating in the circumstellar medium
and  losing its previous degree of symmetry, indicating a very short-lived event.
In addition, our results reveal, at scales of $\sim$ 1$''$ (700 AU), the
simultaneous presence of a relatively slow ($\sim$ 10-70 km~s$^{-1}$) wide-angle outflow 
(opening angle of $\sim$ 102$^{\circ}$), traced by the masers,
and the fast ($\sim$ 500~km~s$^{-1}$) highly collimated radio jet associated with HW2 (opening angle of $\sim$ 18$^{\circ}$), previously observed with the VLA. This simultaneous presence of a wide-angle outflow and a highly collimated jet associated with a massive protostar is similar to what is found in some low-mass YSOs. There are indications that the primary wind(s) from HW2 could be rotating. The implications of these results in the study of the formation of high-mass stars
are discussed.

\end{abstract}


\section{Introduction}

The main characteristics of the formation process and evolutionary
sequence  of low-mass young stellar objects (YSO) is relatively
well understood, with their formation via accretion disks (precursors
of planetary systems) and the simultaneous ejection of collimated
winds through the poles of the disk. The subsequent interaction
of the winds with the ambient medium gives rise to other observed
phenomena, such as molecular outflows and  Herbig-Haro (HH) systems
\citep{a1,s1,b1,m1}. In the case of massive stars (M $\gtrsim$
8~M$_{\odot}$), the progress in the understanding of their formation
and influence on the environment medium has been more difficult
\citep{h1,h2,z1}. 
It might be possible that high-mass stars form in a similar 
way than low-mass stars, namely through accretion onto the protostar 
until a massive object is assembled.
However, there are significant differences between low- and
high-mass star formation, among them: i) Massive stars are
on the main sequence almost from their birth, with radiation pressure
acting on dust grains that can become large enough to reverse the
infall of matter (in the case of low-mass stars, radiation pressure
does not play any significant role in the halting of the accretion process), and
ii) they form in clusters and associations, within warmer, larger
and more massive molecular clouds, with strong and complex influences
on the surrounding molecular material via ionising radiation,
outflows, and gravitational interactions. These differences
have led to the suggestion that different mechanisms (or modes) may
apply to low- and high-mass star formation; for example, through
an accretion disk in a similar way to the low-mass case, but with
very  high accretion rates (Yorke \& Sonnhalter 2002; McKee \& Tan 2003; Krumholz et al. 2005, 2009), through a competitive accretion in a protocluster environment (Bonnell \& Bate 2006), or through coalescence
of previously formed  stars of lower mass \citep{b2}.

One observational approach in the study of high-mass star formation
has been to search for disks and jets in young
high-mass stars, trying to find a parallelism with the formation
of  low-mass stars (see, e.g., the disk-outflow systems associated
with the massive stars in W51N and W33A at scales of $\sim$ 2000--3000~AU; Zapata et al. 2009; Davies et al.
2010; and the radio jets in HH80-81, AFGL 2591, IRAS 20126+4104, and IRAS 18089-1732; Mart\'{\i}, Rodr\'{\i}guez, \& Reipurth 1998; Trinidad et al. 2003; Hofner et al. 2007; Beuther \& Walsh 2008). However, only a handful of sources have been studied in
detail due to several strong observational limitations. These
limitations arise from the fact that there is a relatively small number of
young high-mass stars (they are a very small fraction of the new stars and besides evolve quickly),
they are highly obscured (they are frequently only observable at
cm and [sub]mm wavelengths), and they are remote, which makes very
difficult to identify and isolate single young high-mass stars owing
to the lack of sufficient angular resolution.  For example, an
angular  resolution of 1$''$  implies a linear scale of 2000 AU for
sources at distances of 2 kpc (a typical distance for high-mass
star formation regions),  which are scales where multiple high-mass
objects are expected to be found (extrapolating from what is observed
in Orion, the closest region of massive star formation;  e.g.,
Zinnecker \& Yorke 2007).

At a distance of 725~pc (Johnson 1957; Moscadelli et al. 2009), Cepheus-A is the second
nearest high-mass star formation region after Orion (see Kun, Kiss, \& Balog
2008 for a review of the main observational characteristics of this
region). Because of its relative proximity, it has been possible
to identify and isolate at the centre of this region the young
massive star Cepheus A HW2, the brightest of the radio continuum sources
observed in the region \citep{h3,r1,g1}.  With a mass of $\sim$
15-20~M$_{\odot}$, HW2 is deeply embedded in a high-density core with
visual extinction A$_V$ $\simeq$ 10$^3$ mag (Torrelles et al. 1985,
1993).  In fact, Pravdo et al. (2009) and Schneider, G\"unter, \& Schmitt (2009) have recently found that while some of the other radio continuum sources
near  HW2 are detected in X-rays with Chandra, HW2 itself
is not detected at these frequencies, deriving very high
absorbing column densities towards this massive object (although see Parkin et al. 2009 showing that massive YSOs are hard to detect in X-rays even with moderate column densities).
HW2 appears also hidden at 24.5~$\mu$m \citep{d2}. The nature of
HW2 as a young massive star is also inferred from its association
with very bright masers and intense magnetic fields
\citep{t3,m3,g2,n1,b3,v1,c1,p2,s3}. However, what makes this source
a singular young massive star is its association with an ionised
biconical highly collimated jet at scales  $\lesssim$ 1000~AU (outflow opening angle $\sim$ 18$^{\circ}$), exhibiting ejections
in opposite directions moving away from the central source at $\sim$ 500~km~s$^{-1}$,
with an orientation similar to that of the more extended ($\sim$ 1$'$)
bipolar molecular outflow seen in HCO$^+$  \citep{r1,g3,c2}. These
observational characteristics, in analogy to what is observed in
low-mass YSOs, supports the hypothesis that the massive young star HW2 has
been formed through an accretion disk, rather than by merging of
lower mass stars, because in this last case a collimated jet is not expected to survive. The detection
of a rotating disk of dust and molecular gas with a size of $\sim$
700~AU oriented perpendicular to, and spatially coincident with,
the HW2 radio jet, together with the magnetic field predominantly
aligned along the outflow and perpendicular to the disk, gives
further support to the accretion disk scenario for the formation
of this massive star \citep{p3,j1,t6, v2}.

The presence of a disk--YSO--jet system associated with Cepheus A HW2 (the smallest one ever detected around a young massive star),  which is part of a cluster of massive stars observed in the region
\citep{g1,c2,m4,j2}, has turned this object into an  ideal laboratory for
testing current theories of high-mass star formation, as well as
to study their influence  on the surrounding high-density molecular
material via radiation (e.g., ionisation, dust and molecular gas
heating; Garay et al. 1996; Jim\'enez-Serra et al. 2009), outflows
(e.g., shock-excited molecular gas emission; Hartigan et al. 1986,
2000; Torrelles et al. 2001a, b; Patel et al. 2007), and gravitational
interactions between the members of the cluster (e.g., precession
of the disk-HW2-jet system; Curiel et al. 2006; Cunningham, Moeckel,
\& Bally 2009).

In this paper, we  present Very Long Baseline Interferometry  (VLBI)
multi-epoch water maser observations of the Cepheus A region. The 
high brightness and the compact nature of masers have proven to 
be extremely useful to study with angular resolution better than 1 mas the 
main properties of the gas very close to the YSOs. The
main goal of  the observations presented here was to follow the evolution of the
remarkable water maser linear/arcuate structures found in earlier
VLBI observations \citep{t4,g2}. These previous observations led
to the identification of at least three different centres of star formation
activity in a region of $\sim$ 0.3$''$  (200~AU) radius; the first one
associated with HW2, a second one associated with the exciting
source of  the expanding water maser bubble R5, and the third one
associated with the exciting source of the maser arcuate structure
R4. Our new data  were obtained from five epochs of observation (\S 2). In \S~3 we present our main observational
results, and their implications in the study of the formation of high-mass
stars are discussed in \S 4. The main conclusions of this work
are presented in \S 5.

\section{Observations}

The multi-epoch water maser line (6$_{16}$$\rightarrow$5$_{23}$; 
$\nu$ = 22235.080 MHz) observations
were carried out with the Very Long Baseline Array (VLBA) of the
National Radio Astronomy Observatory  (NRAO)\footnote{The NRAO is
a facility of the National Science Foundation operated under
co-operative agreement by Associated Universities Inc.} on 2001 July
11, July 30, August 18, September 13, and 2002 January 27, for
$\sim$ 11 hours during each of the five epochs. We used 512 spectral
line channels of  0.21 km~s$^{-1}$ width, covering a velocity range
$V_{\rm LSR}$ $\simeq$ $-$66 to +42~km~s$^{-1}$. After correlation
at the NRAO Array Operation Center (AOC), the data were calibrated
and imaged using the Astronomical Image Processing System (AIPS).
Delay and phase calibration was provided by 3C345, 1739+522, 2007+777,
BL Lac, and 3C454.3, while bandpass corrections were made using
3C345, BL Lac, and 3C454.3. The synthesised beam size  was $\sim$
0.4~mas for the five epochs. We identified a strong ($\geq$ 10~Jy)
point-like water maser at  $V_{\rm LSR}$ = $-$6~km~s$^{-1}$ persisting
in all the observed epochs,  allowing us to self-calibrate the data using
that same reference feature. The absolute co-ordinates obtained for
the reference feature in the five epochs were in agreement to within
$\sim$ 0.01$''$.  
A first co-ordinate alignment of the five epochs
of observations was made with respect to this reference feature.

In order to identify the regions of  water maser emission, 
we first made large data cubes of
512 channel images of 8192$\times$8192 pixels with a cell-size of 1 mas for each
of the five epochs,
centred on the reference feature that was used for self-calibration.
Water maser emission was found essentially around HW2 and HW3d,
spread over a spatio-kinematical region of $\sim$ 4$''$ and
$\sim$ 200 velocity channels ($V_{\rm LSR}$ $\simeq$ $-$30 to
+10~km~s$^{-1}$). After that, we simultaneously mapped, for each
epoch, 7 fields of 4096$\times$4096 pixels $\times$ [200 velocity
channels] with a cell-size of 0.1 mas to image, with the full angular
resolution, the space-velocity regions identified with lower angular
resolution in the previous step. In this way, about 1800 maser spots
for each epoch were identified, most of them unresolved, obtaining their positions by
fitting a two-dimensional Gaussian profile (we define a maser spot as emission
occurring at a given velocity channel and distinct spatial
position). The rms noise level of the
individual channel maps ranges from $\sim$ 5 to 400 mJy~beam$^{-1}$,
depending on the intensity of the maser components present in each velocity
channel. For the identification of the maser spots a signal-to-noise
(S/N) ratio $\ga$ 8 was adopted as the detection threshold. From
this S/N ratio and the beam size of 0.4 mas, we estimate that the accuracy
in the relative positions of the spots within each epoch is better
than 0.03 mas.

In general, VLBA proper motion measurements of water masers are
limited by the high time variability of their flux density. The
standard method is to phase reference to a particular strong maser
spot, as the one used here for self-calibration (see above). However,
the selected reference spot may have its own proper motion, which
therefore results in arbitrary offsets in the proper motions for
the whole system. For the measurement of proper motions of water masers
presented in this paper (\S 3), we defined a more ``stationary" reference
frame, by identifying nine strong masers ($\ga$ 1
Jy) with different LSR velocities, spread over the full region of
$\sim$ 4$''$ size where maser emission is detected, and persisting
in the five observed epochs. The new reference position is defined
as the mean position of these nine strong masers. We calculated this mean position for each epoch, and assuming
that it is stationary, we realigned all epochs with respect to it. After that, all the
offset positions presented in this paper are with respect to the nominal position of HW2 ($\alpha$(J2000.0) = {\rm $22^h56^m17.982^{s}$}, $\delta$(J2000.0)
= 62$^{\circ}$01$'$49.57$''$;  C2006). In the following
section we present the main results obtained toward HW2 and HW3d.
All the proper motion values given in this paper have been determined 
considering the first four epochs, which are grouped in a short
time-span of two months, thus minimising 
variability effects.
The fifth epoch, separated from the fourth one by four and a half months, 
has been  used to confirm trends in the derived proper
motions, but not in their explicit calculation.

\section{Results and discussion}

In Figure 1 we show the positions and radial velocities (LSR) of the water maser
spots detected with the VLBA in the five observed epochs (hereafter
VLBA 2001-2002 data set). In this figure we also show the 1.3 cm
continuum and water maser sources observed previously with the VLA
(hereafter VLA 1995 data set; T98), as well as the water masers observed
in 1996 with the VLBA (hereafter VLBA 1996 data set; T2001a, b). The
alignment of the different data sets (VLA 1995, VLBA 1996, and VLBA 2001-2002) has been made by their absolute
co-ordinates. The
accuracy of this simple alignment is better than 0.05$''$, which
is good enough for a general comparison between the different data
sets at the scale shown in Figure 1. In this figure we have also
labelled several sub-regions ``R'' (using the naming convention of
T2001a, b and Gallimore et al.  2003). We can see that the
spatio-kinematical distribution of the masers around HW2 and HW3d
traced by the different data sets coincides quite well at scales
$\ga$ 1$''$ ($\ga$ 700 AU), without major changes over the 1995-2002
period. This implies that both the surrounding interstellar medium
and the exciting sources of the water masers have been relatively
stable over these spatial and temporal scales. 

\begin{figure}
 \includegraphics[scale=0.83]{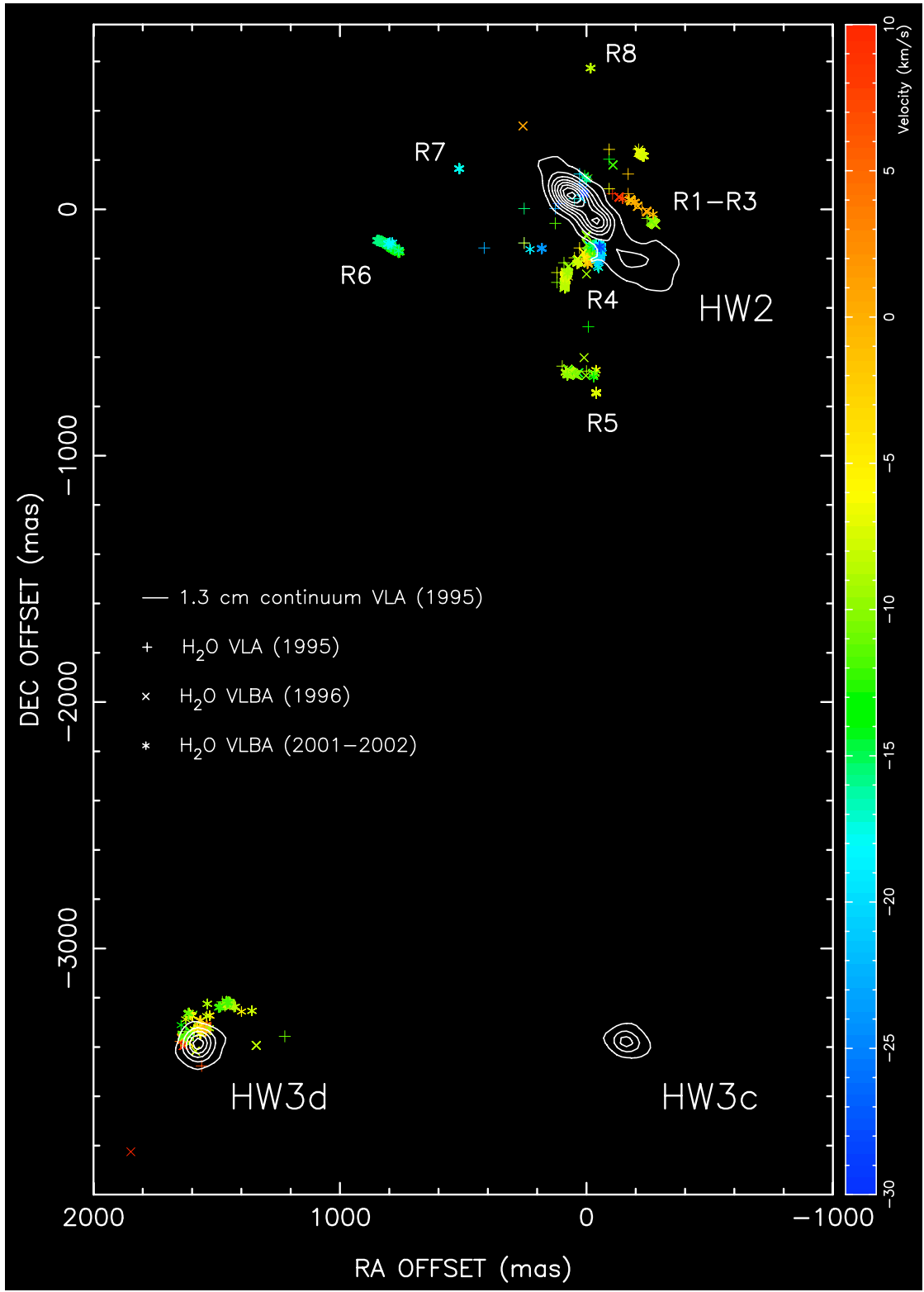} 
 \caption{Positions of the water maser spots measured with the VLA 1995 (beam size $\simeq$ 80 mas; T98),
VLBA  1996 (beam size $\simeq$ 0.5~mas; T2001a,b), and VLBA 2001-2002 (beam size $\simeq$ 0.4~mas; this paper) data sets, overlaid onto the 1.3 cm continuum contour map showing the HW2 radio jet, and the HW3c, and HW3d radio sources (beam size $\simeq$ 80~mas; T98). Colour code indicates the LSR radial velocity (km~s$^{-1}$) of the individual maser spots. Different sub-regions ``R'' discussed in this paper are numbered (see \S\S~3.1, 3.2, and 4, and Figs. 2, 3, and 4). Offsets positions are referred with respect to the HW2 position, $\alpha$(J2000.0) = {\rm $22^h56^m17.982^{s}$}, $\delta$(J2000.0)
= 62$^{\circ}$01$'$49.57$''$ (C2006).}
\end{figure}

The water masers associated with HW3d are mainly located at the
border of the radio continuum emission as seen in the two VLBA data
sets (1996 and 2001-2002; Fig. 1). These masers cover a velocity
range V$_{LSR}$ $\simeq$ --16 to +8 km~s$^{-1}$, with intensities from
$\sim$ 0.03 to 130 Jy~beam$^{-1}$ (brightness temperatures up to
$\sim$ 2$\times$10$^{12}$~K), and showing a more compact distribution
(within $\sim$ 0.1$''$ radius, $\sim$ 70 AU) than the
masers  around HW2 (Fig. 1). Contrary to the masers around HW2, the
masers spatially associated with HW3d show no evidence of any 
clear linear/arcuate structure, any velocity segregation as a function 
of position, nor clear proper motions. The association of intense water masers
with HW3d further supports that this source harbours an internal young
massive star (early B star) as suggested by \citet{g1}, although
perhaps less evolved than HW2 because of the compactness of both
the 1.3 cm continuum emission and the cluster of masers.

With regard to the masers around HW2, large changes are observed
at scales $\la$ 0.1$''$ (70 AU) in the linear/arcuate water maser
structures found in the earlier VLBA 1996 data,  including the
disappearance of the R1 linear structure in the VLBA 2001-2002
data, and the evolutionary changes of the R4 and R5 expanding
arcuate structures. In addition, we detected new maser structures
labelled as R6, R7, and R8 in Figure 1 that were not seen in
the previous VLBA 1996 data set.  In what follows (\S\S 3.1, 3.2),
we present and discuss all these results obtained with the VLBA
2001-2002 data set, analysing the spatio-kinematical distribution
of the masers in the different sub-regions, including proper motion
measurements.

\subsection{Spatio-kinematical distribution of the R6, R7, and R8 maser structures}

{\bf R6}. This remarkable linear structure, with a velocity range V$_{LSR}$ $\simeq$ --18 to 
--11~km~s$^{-1}$ and intensities from
$\sim$ 0.03 to 14 Jy~beam$^{-1}$, is located  $\sim$ 0.8$''$ (580 AU) east of HW2 (Fig. 2), and  persists as a whole in the five epochs 
of the VLBA 2001-2002 data set (R6 was also detected by Vlemmings et al. 2006 in their single 
epoch observations carried out in 2004 with the VLBA).  This linear  structure (Fig. 2) is 
oriented Northeast-Southwest (PA $\simeq$ 66$^{\circ}$), has  a size of $\sim$ 100~mas in length (this 
is, $\sim$ 250 times larger than the VLBA beam), and is formed by 2-3 roughly 
parallel ``layers" in each one of the individual observed epochs, with the 
``upper layer" blue-shifted (V$_{LSR}$ $\simeq$ --16 km~s$^{-1}$) 
with respect to the ``lower layer" (V$_{LSR}$ $\simeq$ --13 km~s$^{-1}$). In addition, 
a LSR radial velocity gradient of $\sim$ 4-5~km~s$^{-1}$ over a scale of 0.1 arcsec ($\sim$ 
0.06~km~s$^{-1}$~AU$^{-1}$) is observed along the linear structure, with more 
red-shifted velocities toward  the edge closer to HW2, and more blue-shifted velocities outward (this is 
more clearly seen in the ``lower  layer"; Fig. 2). The entire linear structure
shows a transversal motion in the sky, with a proper motion of 
$\sim$ 13~km~s$^{-1}$ toward the Southeast  (PA $\simeq$ 118$^{\circ}$; see 
Fig. 2 and Animation in the supporting information section).

{\bf R7}. This structure is detected in the first four epochs of the VLBA 2001-2002 data set at $\sim$ 0.6$''$ (400 AU) Northeast of HW2. It is located in between R6 and R8 (Figs. 1 and 2), forming a linear structure oriented Northwest-Southeast (Fig. 2), with the
masers having radial velocities in the range V$_{LSR}$ $\simeq$ --18 to --16~km~s$^{-1}$ and intensities from
$\sim$ 0.03 to 0.25~Jy~beam$^{-1}$.
Interestingly, while keeping both its radial velocity distribution and linear structure, it increases progressively its linear size as a function of time, from 1.3 mas (0.9 AU)  to  1.7 mas (1.2 AU), 1.9 mas (1.4 AU), and 2.1 mas (1.5 AU), for the first, second, third, and fourth epoch, respectively. There is also  a clear LSR radial velocity gradient of $\sim$ 1~km~s$^{-1}$~mas$^{-1}$ (1.4~km~s$^{-1}$~AU$^{-1}$) along the linear structure seen in all the observed epochs, with more red-shifted velocities at the north-western parts (Fig. 2). Furthermore, the entire R7 linear structure is moving in the sky toward the north-east (PA $\simeq$ 74$^{\circ}$; i.e., nearly perpendicular to the direction of the linear structure), forming a relatively fast opening outflow with proper motions of $\sim$ 70~km~s$^{-1}$ (see Fig. 2 and Animation in the supporting information section). These motions are a factor $\sim$ 5 larger than the proper motions observed in R6 and R8 (see below) and present a significantly different direction. On the other hand, R7 moves with a proper motion $\sim$ 7 times smaller than
those observed in the ionised jet with PA $\simeq$ 45$^{\circ}$ (C2006).

\begin{figure}
 \includegraphics[scale=.6]{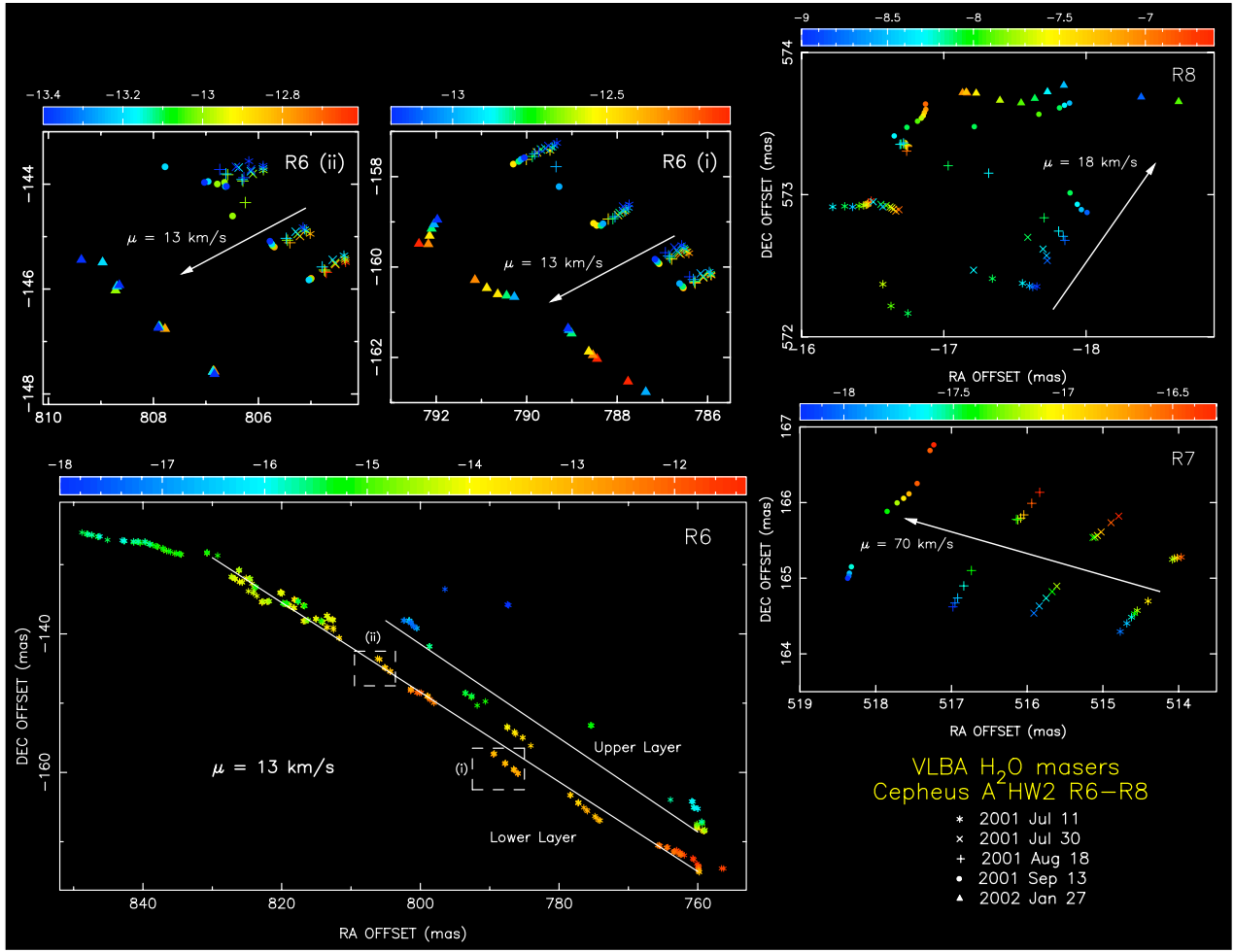} 
 \caption{Positions of the water maser spots measured
in sub-regions R6, R7, and R8 in the five epochs of the VLBA 2001-2002 data set (see also Fig. 1). These masers were not detected in the VLBA 1996 data set (T2001b). Each of the five epochs is represented by a different symbol. Colour code indicates the LSR radial velocity (km~s$^{-1}$) of the individual spots. 
Proper motions are clearly seen in all these sub-regions (solid arrows indicates the mean values and directions of these motions in the sky). For the structure R6 (bottom left), we have only drawn the positions measured in the first epoch to show more clearly its linear structure and the two parallel layers discussed in the text, although close-ups of two small sections including the five epochs are also shown. See also Animations in the supporting information showing the motions of the masers in the sky.}
 \end{figure}
 
{\bf R8}. It is located $\sim$ 0.6$''$ (400 AU) north of HW2, with their masers having radial velocities in the range V$_{LSR}$ $\simeq$ --9 to --7 km~s$^{-1}$ (intensities from
$\sim$ 0.05 to 5~Jy~beam$^{-1}$) persisting in the five observed epochs. In this region, the masers show no clear linear/arcuate structures (except for the linear structure observed in epoch 3, Fig. 2), nor a clear radial  velocity gradient. However, maser motions in the sky are clearly seen  toward the north-west, with values of $\sim$ 18~km~s$^{-1}$ (PA $\simeq$ --35$^{\circ}$; Fig. 2 and Animation  in the supporting information section). These proper motion values, while significantly smaller than those measured in R7, are of the order of  those measured in R6 but with different orientation (Fig. 2).

The general properties of the masers in R6-R7-R8 can be explained within a shock interaction
scenario within the disk-HW2-outflow system, with the simultaneous
presence of a fast highly collimated jet and a slower wide-angle outflow associated
with the massive  protostar HW2 (see below, \S 4). Masers in the R6-R7-R8 structures appear
globally blue-shifted (V$_{LSR}$ $\simeq$ --18 to --7 km~s$^{-1}$) with respect to the velocity system
of the circumstellar disk associated with HW2 (V$_{LSR}$ $\simeq$ --7 to --3 km~s$^{-1}$; Patel et al. 2005, Jim\'enez-Serra et al. 2007, 2009). We attribute this global blue-shift of the R6-R7-R8 masers
to the fact that they are located on the blue-shifted lobe of the more extended ($\sim$ 1$'$) Northeast-Southwest
bipolar molecular (HCO$^+$) HW2 outflow (G\'omez et al. 1999), and, therefore,  they present an additional (global) blue-shifted component also moving
towards the observer.

\subsection{Evolution of the R5 and R4 arcuate maser structures}

{\bf R5}. This sub-region contains the arc structure detected by
T2001a, b with the VLBA 1996 data set.  In those data, the R5 arc structure, with a size of
$\sim$ 100 mas (700 AU), defined a circle of radius $\sim$ 60 AU
with an accuracy of one part in a thousand. It expanded
at $\sim$ 9~km~s$^{-1}$, and was interpreted by these authors as
caused by a short-lived episodic spherical ejection of material
(dynamical time-scale of $\sim$ 30 yr) from a massive young star
located at the  centre of the fitted circle. In fact, after the
recognition of this maser structure, Curiel et al. (2002) detect
a very weak 3.6 cm-continuum source ($\sim$ 0.2 mJy) toward the
 centre of the circle ($\sim$ 0.6$''$ south of HW2), and
they propose it as candidate for the excitation of the R5 masers.
However, the nature of this faint cm-continuum source is still unknown
(e.g., we do not know its spectral index yet).

\begin{figure}
 \includegraphics[scale=0.75]{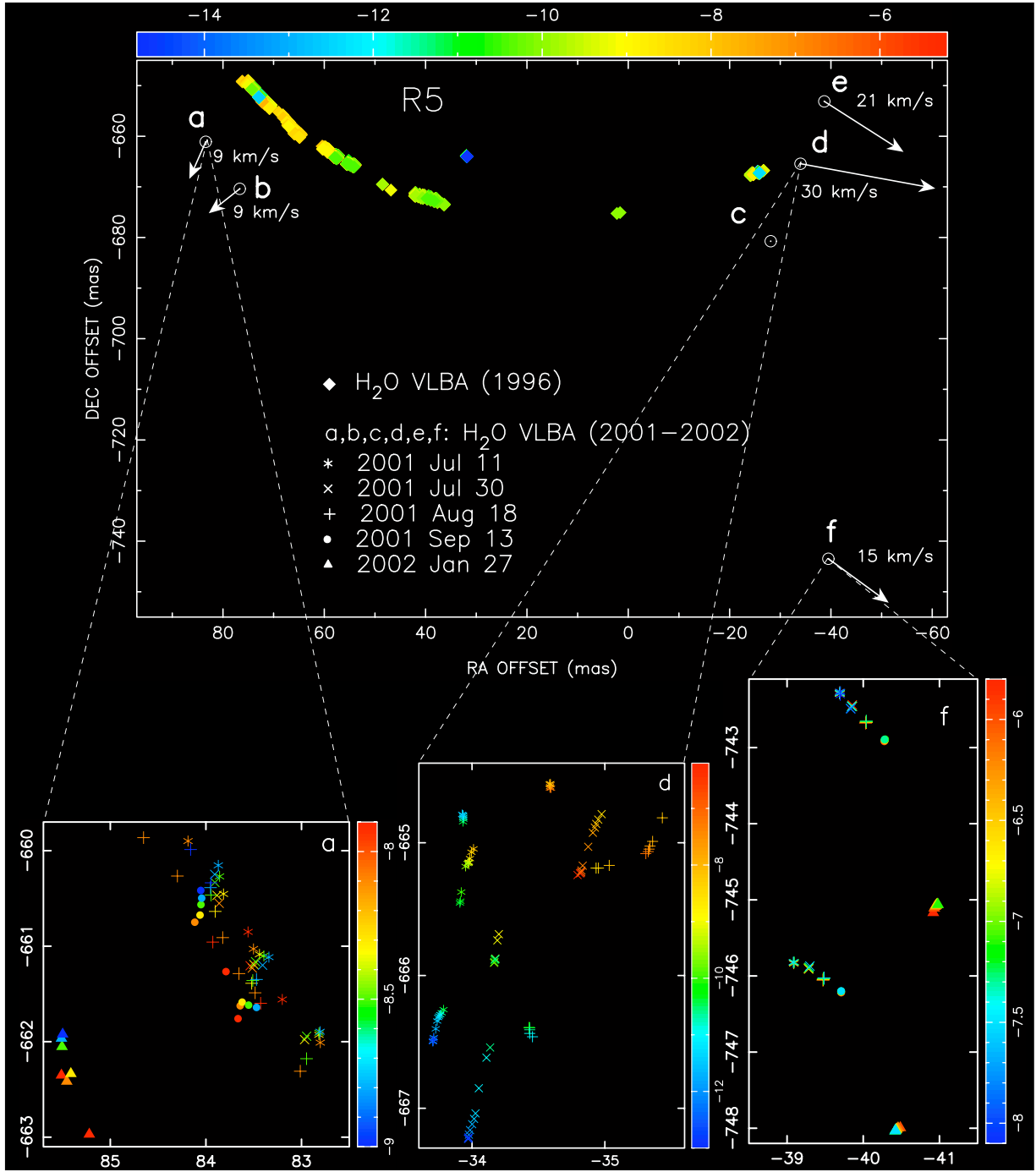} 
 \caption{Spot positions measured
in the five epochs of the VLBA 2001--2002 data set in sub-region R5 (see Fig. 1).  Six groups of masers are detected (a, b, c, d, e, f), some of them (a, d, f) shown as a close-up. Colour code indicates the LSR radial velocity (km~s$^{-1}$) of the spots. The R5 arc structure observed with the VLBA 1996 data (T2001a,b) is also shown in the upper panel. Vectors indicate the direction and values of the mean proper motions measured in the a, b, d, e, and f maser groups (proper motions were not measured in c). See Animation  in the supporting information showing the motions in the sky of these maser groups.}
 \end{figure}

In Figure 3, we show the spatio-kinematical distribution of the
masers as observed with the VLBA 2001-2002 data set in R5. A
co-ordinate alignment with the VLBA 1996 data set was made assuming
that the strongest maser within the R5a group (VLBA 2001-2002 data set; Fig. 3)
corresponds to the strongest maser within the group labelled as R5-V
(VLBA 1996 data set; Figure 6 of T2001b), both having similar velocities
(V$_{LSR}$ $\simeq$ --8.5~km~s$^{-1}$). For a proper alignment, we estimated the position
that the strong maser in
R5-V (VLBA 1996 data set) should have had in the  VLBA 2001-2002 data set, assuming that
the proper motions of the expanding structure (9 km~s$^{-1}$) remained
constant over the whole time period. The final alignment resulted in
RA and Dec shifts on the VLBA 1996 data set of --25 and --10 mas, respectively. The value of this  shift is
smaller than the estimated error of  the simple alignment by the
absolute co-ordinates ($\sim$ 50 mas). We point out that this
alignment is only valid for general description/comparison purposes
between the different VLBA 1996 and 2001-2002 data sets, but not
for detailed studies. This caution arises because
we cannot be sure that the individual maser spot selected for the alignment
is the same maser spot that has persisted during the large time span of five years.

With the new VLBA 2001-2002 data set, we note that the R5 maser
structure has persisted during five years, although its spatial
morphology has significantly changed over this time span. 
In these new data we identify six 
regions that we label as R5a--f (Fig. 3). Regions R5a, R5b, and R5c 
apparently correspond to the expansion of the ring, while most of the arc 
structure observed in 1996 is not present in the 2001-2002 data.
In addition,  new ``subgroups" of
masers have appeared, labelled in Figure 3 as R5d and R5e to the
west, and R5f to the south-west.  Interestingly, the maser emission
detected with the VLBA 2001-2002 data set covers approximately the
same radial velocity range as in 1996 (V$_{LSR}$ $\simeq$  --15 to --
5~km~s$^{-1}$; S$_{\nu}$ $\simeq$ 0.03--16~Jy~beam$^{-1}$),  and  the directions
of the proper motions are consistent
with those previously observed. In fact, while the groups of masers
in the eastern part of the R5 structure (R5a and 5b; Figure 3) are moving
toward the south-east, with proper motion values of $\sim$
9~km~s$^{-1}$,  the group of masers to the west (R5c, 5d, 5e, and
5f) are moving toward the south-west, with proper motion values of
$\sim$ 15-30~km~s$^{-1}$. These properties led us to conclude
that we are indeed observing the same expanding structure as that
reported by T2001a, b, but that this structure is undergoing a distortion 
in its expansion through the circumstellar medium, losing its previous degree of symmetry.

Different scenarios have been proposed to explain the R5 structure,
among them, the mentioned short-lived episodic spherical ejection
of material from a YSO (which is not predicted by current star
formation theories), and alternatively, that the maser structure
corresponds to a shocked layer of ambient molecular material around
an expanding HII region (see Curiel et al. 2002, Lizano 2008). In
both scenarios, the spherical molecular shell evolution could be
affected by internal instabilities and by ambient inhomogeneities
during its expansion, as observed in R5.  In any case, the dissipation
of the R5 structure in the ambient medium corroborates the short-lived
nature of this phenomenon.  

\begin{figure}
 \includegraphics[scale=.75]{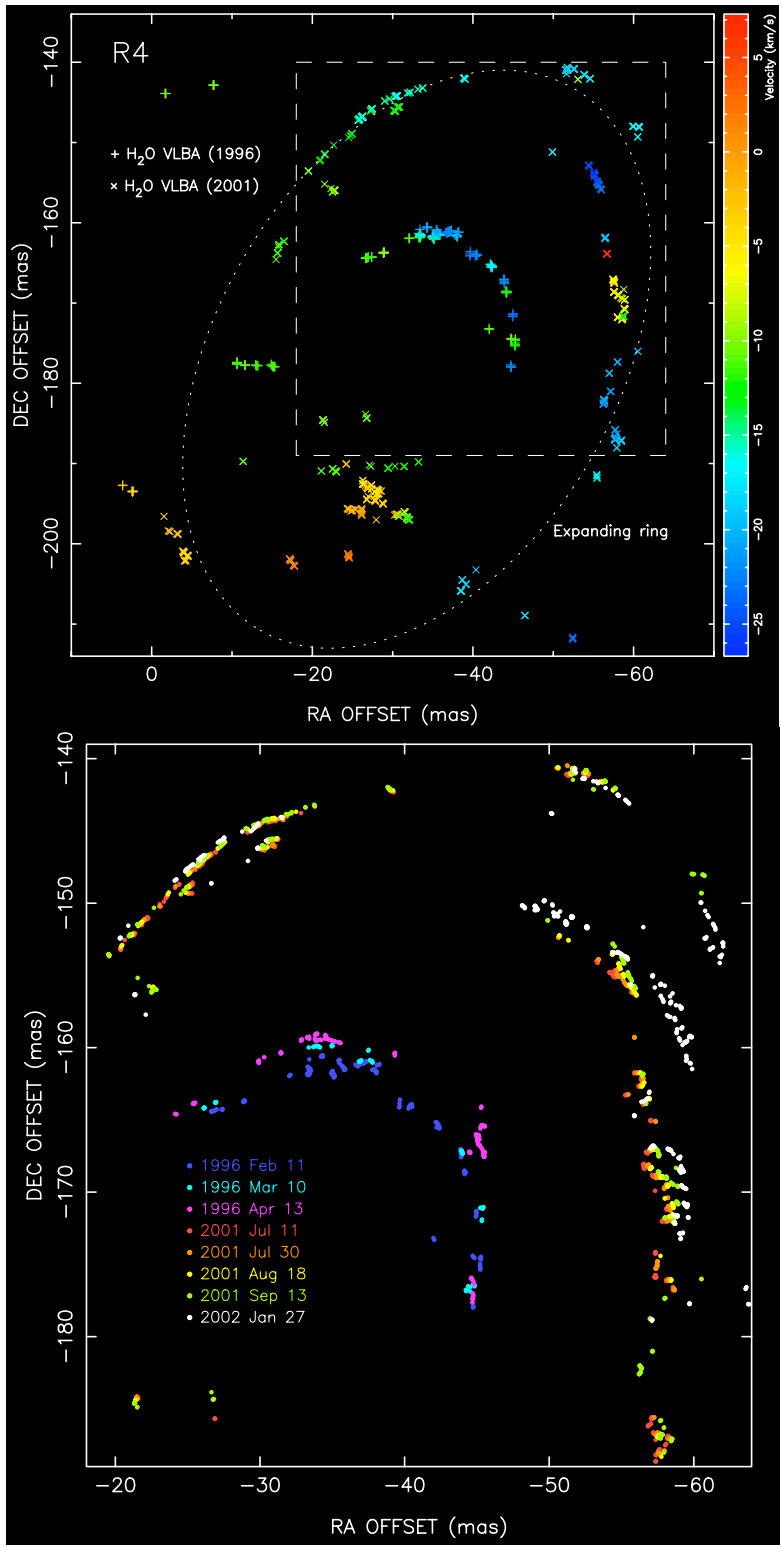} 
 \caption{{\it Upper panel}: Spot positions measured with the VLBA 1996 (T2001b) and 2001-2002 (this paper) in sub-region R4 (see also Fig. 1). Colour code indicates the LSR radial velocity (km~s$^{-1}$) of the spots.
 In this panel only the epochs of 1996 Feb 11 and 2001 Sep 13 are plotted. The VLBA 2001-2002 masers trace in the sky an expanding elliptical ring (dotted line).  {\it Lower panel:} Spot positions measured in all the observed VLBA epochs
(represented by dots with different colours), showing progressively the expanding motions in the sky. From 1996 to 2001-2002, the R4 structure
has moved $\sim$ 25 mas, which represents proper motions  of $\sim$ 5~mas~yr$^{-1}$ (15~km~s$^{-1}$). The alignment of the 1996 and 2001-2002 data sets is the same than the alignment explained in \S 3.2 to plot the masers in Fig. 3. See also Animation  showing the motions in the sky.}
\end{figure}

{\bf R4}. A remarkable arcuate structure of $\sim$ 35 mas size
(25 AU) with proper motions of 10-30 km~s$^{-1}$ was  observed  in
this sub-region with the VLBA 1996 data set ($\sim$ 0.17$''$
south of HW2; Fig. 1). This arcuate structure
had the highest LSR radial velocity dispersion of all the observed
maser structures in Cepheus A, and was interpreted as a bow-shock
moving toward the north-west excited by the wind of an energy source
other than HW2 (T2001b).  Gallimore et al. (2003), through Multi-Element
Radio-Linked Interferometer Network (MERLIN) water maser observations
on 2000 April 9,  also detect this structure, and find a
displacement of $\sim$ 16 mas toward the north-west between the
arc observed in 1996 (VLBA)  and 2000 (MERLIN), corresponding to
proper motions of  $\sim$ 4~mas~yr$^{-1}$ (14 km~s$^{-1}$).
Furthermore, these authors find that the displacement of the arc
is compatible with expansion about a common centre, proposing that
the R4 arcuate structure is part of an expanding elliptical ring (rather
than a simple bow-shock) driven by a wind from a still unidentified
central massive YSO. In this scenario, the elliptical ring would be produced
by a shock wave propagating through a rotating disk surrounding the
YSO.

{\bf The} VLBA 2001-2002 water maser observations show that the R4 arcuate
structure is persisting five years after its first detection,  with
the emission covering a line- of-sight velocity range V$_{LSR}$
$\simeq$ --27 to +7 km~s$^{-1}$ and with a range of intensities $\sim$
0.03-100~Jy~beam$^{-1}$ (among the most intense of the region, together with R5). In Figure 4 we plot together the  VLBA
1996 and 2001-2002 maser spots using the same alignment explained
above for R5. The VLBA 2001-2002 masers trace a nearly elliptical
patchy ring of $\sim$ 70 mas size (50 AU; PA $\simeq$
--30$^{\circ}$), with most of the masers detected in the previous
epochs of the VLBA 1996 data set located inside the  ring (see upper
panel of Figure 4). This ring, which was only partially visible in the VLBA
1996 data set (T2001b), is similar in size and orientation to that
found by Gallimore et al. (2003) with MERLIN (epoch 2000; see their
Figure 7). The most clearly defined part of the ring is the north-western  R4 arcuate structure, which in five years has moved $\sim$
25 mas toward the north-west. This corresponds to a proper motion
of $\sim$ 15 km~s$^{-1}$, consistent with the values measured with the VLBA in the 1996 epoch ($\sim$10-30~km~s$^{-1}$) as well
as with the value measured by Gallimore et al. (2003) from VLBA
1996 to MERLIN 2000 ($\sim$ 14 km~s$^{-1}$). These proper motions are better appreciated in the lower panel of Figure 4, where
we show a close-up of the arc structure. A
progressive spatial displacement is clearly distinguished,
both in the three epochs of the VLBA 1996 observations (T2001b), as well
as in the five epochs of our new VLBA 2001-2002 observations. With our
present data we cannot establish if the expanding motions
have been accelerating or decelerating during this time span of five
years. More recent VLBA water maser observations carried out in 2004
by Vlemmings et al. (2006) suggest  that the ring structure (with much less detected
spots) has suffered
a strong deceleration eight years after its first detection, probably due to mass loading as matter is swept up during
the expanding shock-wave. However, the morphology of R4 is complex, showing several
``shells" in our 2001-2002 data (see Figure 4) and we cannot rule out that
we are seeing different features produced by multiple ejections in the different epochs. 
From the size of the ring ($\sim$ 50 AU), and assuming a constant expansion velocity of  $\sim$ 15~km~s$^{-1}$, we estimate a dynamical time-scale of $\sim$ 8 yr for R4.

Our new VLBA water maser observations give better support to the
expanding ring scenario driven by a central YSO, as  proposed
by Gallimore et al. (2003), rather than supporting a simple bow-shock
scenario (T2001b). It is important to point out that the method of alignment
between the different data sets used here (see above) has been different
and independent of the method used by Gallimore et al. (2003)
to compare the 1996 (VLBA) and 2000 (MERLIN) data sets (these authors
assumed that the proper motion of the expansion  centre of the R4 ring
is negligible), but with very similar final results, reinforcing the idea that
the expanding motions have been originated truly at the centre of the ring.  
With our
data we estimate that the central YSO driving the expanding motions
of the  R4 structure should be located at a position offset by
(--0.03$''$,--0.18$''$) from HW2. 

Given the patchy ring structure, with the masers to the
 Northwest showing a more regular distribution than the ones to the Southeast (Fig. 4),  an alternative explanation for the R4 structure
could be that this is an expanding
reverse shock where an external wind (from HW2) is interacting with a small
dense clump located at the centre of the structure (reverse shocks have been predicted by Parkin et al. 2009 within wind-cavity interactions in massive YSOs). However, although we cannot discard this possibility,
the fact that the masers in R4 show the highest LSR radial velocity dispersion of all the maser structures in the region, well differentiated from the others, and with very high flux densities, led us to favour that they have been excited by a close massive YSO other than HW2 through some kind of violent ejection event.

According to the current literature,
we did not find  any known source at the centre of the R4 structure. The
detection of this source is mandatory to ascertain its nature and test
the expanding ring scenario. We think that the future e-MERLIN and
EVLA instruments, with a continuum sensitivity improvement over the
VLA by a factor of $\sim$ 10, will be well suited to reach this goal.

\section{Jet and wide-angle outflows in HW2}

Since the first bipolar molecular outflows and highly collimated
jets from YSOs were discovered (Snell et al. 1980; Rodr\'{\i}guez
et al. 1980; Mundt \& Fried 1983), one of the main issues
that immediately arose was to know their driving and collimating
mechanism, as well as to ascertain the relationship between molecular
outflows and jets. Multiple observations toward low-mass protostars
show than when highly collimated high-velocity jets are
detected (usually in visible or radio continuum wavelengths), they
are enclosed within relatively low-velocity
outflows with a wide opening angle (usually observed in molecular line transitions). Some
clear examples of the simultaneous presence of a jet and a wide-angle
outflow in low-mass YSOs can be found in  L1551 (Itoh et al. 2000),
HH 46/47 (Velusamy, Langer, \& Marsh 2007), HH 211 (Hirano et al.
2006), and IRAS 04166+2706 (Santiago-Garc\'{\i}a et al. 2009); see
also Machida, Inutsuka, \& Matsumoto (2008), and references therein.
Various theoretical models have been developed to explain these
outflows and their relationship in low-mass YSOs, archetypical among them
are the ``X-wind'' and the ``Disk-wind'' models. In the X-wind model, the
primary wind is driven magnetocentrifugally from the interface
between the magnetosphere of the YSO and the associated circumstellar
disk. The primary wind has an angle-dependent density
distribution, with a dense axial jet surrounded by a more tenuous
wide-angle wind. The morphology of both the observed highly collimated jets and the wide-angle molecular outflows are explained through the interaction of the primary wind with the ambient gas 
(e.g., Shu et al. 1994; Shang et al. 2006). On the other hand, 
in the Disk-wind model the highly collimated wind is 
driven magnetocentrifugally
from a wide range of circumstellar disk radii, surrounded by a
wide-angle wind driven by toroidal magnetic pressure during the
evolution of a rotating collapsing magnetised molecular core (e.g.,
Banerjee \& Pudritz 2006). 
Recently, Machida et al. (2008) follow, through MHD simulations,
the evolution of a collapsing cloud,
until protostar formation  takes place. These authors find two distinct flows with
different degrees of collimation and velocities: a low-velocity
wide-angle outflow, magnetocentrifugally driven from the circumstellar disk,
and guided by hourglass-like field lines, and a fast highly-collimated
outflow driven by magnetic pressure and guided by straight field
lines near the protostar. The main difference in terms of observational
predictions between the Machida et al. and the other models is
that in the former, both high-velocity and low-velocity outflows are
naturally explained.  All these
models are very valuable as a first step to understand both the
driving agent of the primary wind(s) and the simultaneous presence
of highly-collimated jets surrounded by wide-angle outflows observed
in low-mass YSOs. Current observations, however,  do not
have enough angular resolution and sensitivity to resolve the
launching region of the primary wind(s) at scales of a few AU, and
therefore it is not possible to discriminate between the different
models proposed (see the review by Ray et al. 2006).

\begin{figure}
 \includegraphics[scale=0.93]{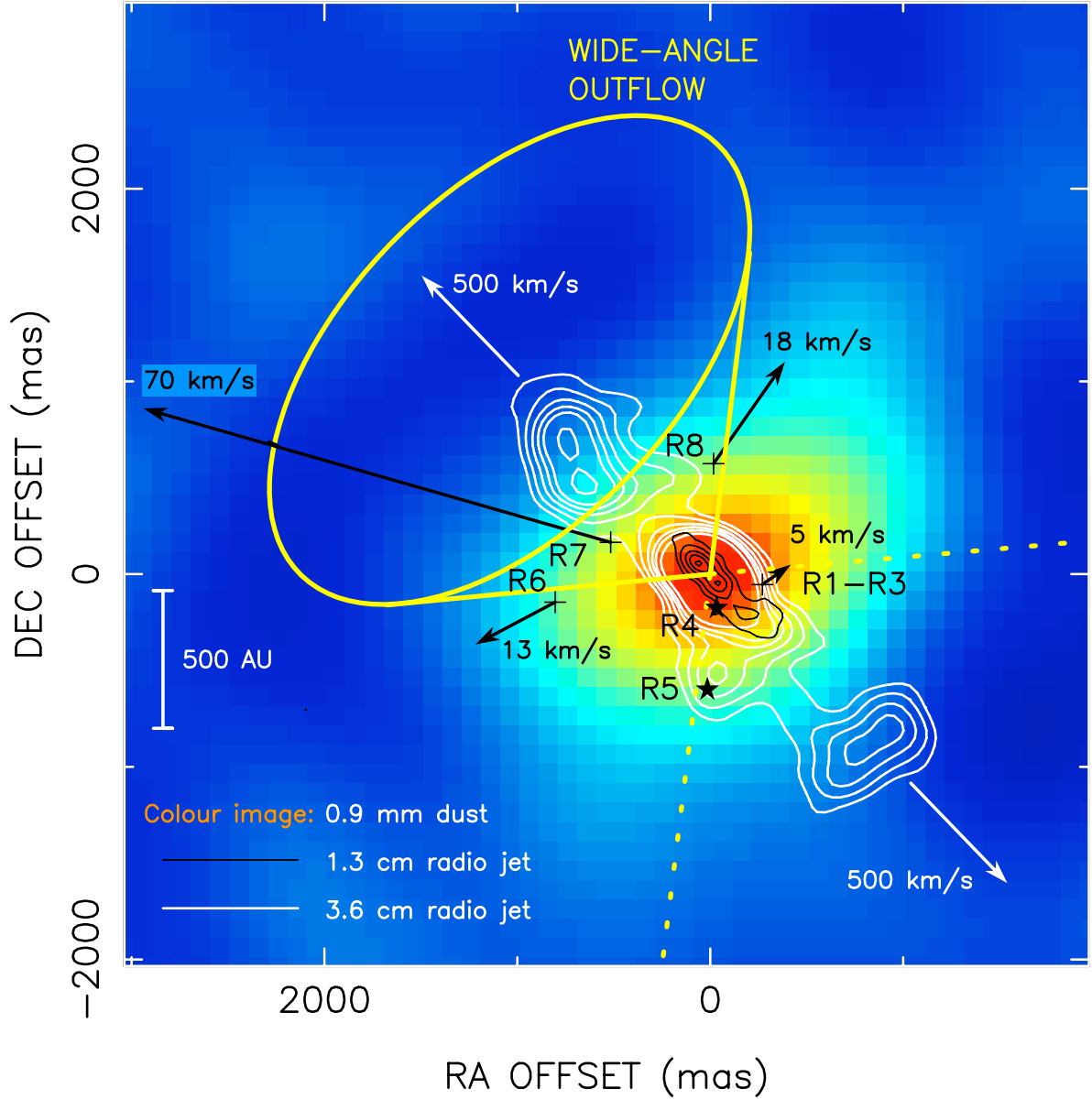} 
 \caption{Wide-angle outflow and jet scenario in HW2. The dust continuum disk of HW2 (Patel et al. 2005) is shown in this figure as a colour image superposed to the highly collimated radio jet (T96, C2006). The radio jet, with an outflow opening angle of $\sim$ 18$^{\circ}$, exhibits ejections in opposite directions moving away at $\sim$ 500~km~s$^{-1}$ from the central source (C2006). R6, R8 and R1-3 trace emission fronts from the walls of the expanding cavities, created and shocked by the wide-angle primary wind of HW2 (opening angle $\sim$ 102$^{\circ}$). The R7 masers, with motions along an axis at an angle
 of $\sim$ 30$^{\circ}$ with respect to the radio jet axis,  are excited inside the cavity by a wide-angle primary wind. They exhibit larger motions than R6, R8, and R1 located at the expanding cavity walls, but lower than the motions of the radio jet. The R6, R7, and R8 masers are blue-shifted with respect to the systemic velocity of the circumstellar disk, while R1-3 are red-shifted. The north-eastern part of the wide-angle outflow is directed toward us, as indicated by
the blue-shifted motions observed in that direction of the more extended ($\sim$ 1$'$) bipolar molecular outflow (G\'omez et al. 1999) and the R6-R7-R8 water maser motions. The position of the two nearby massive YSOs required to excite the R4 and R5 arcuate maser structures are indicated by star symbols (see text). The star associated with R4 is not yet detected. Note that the arrows indicating the motions of the radio jet are shown much smaller than those of the water masers.}
\end{figure}

\begin{figure}
 \includegraphics[scale=.66]{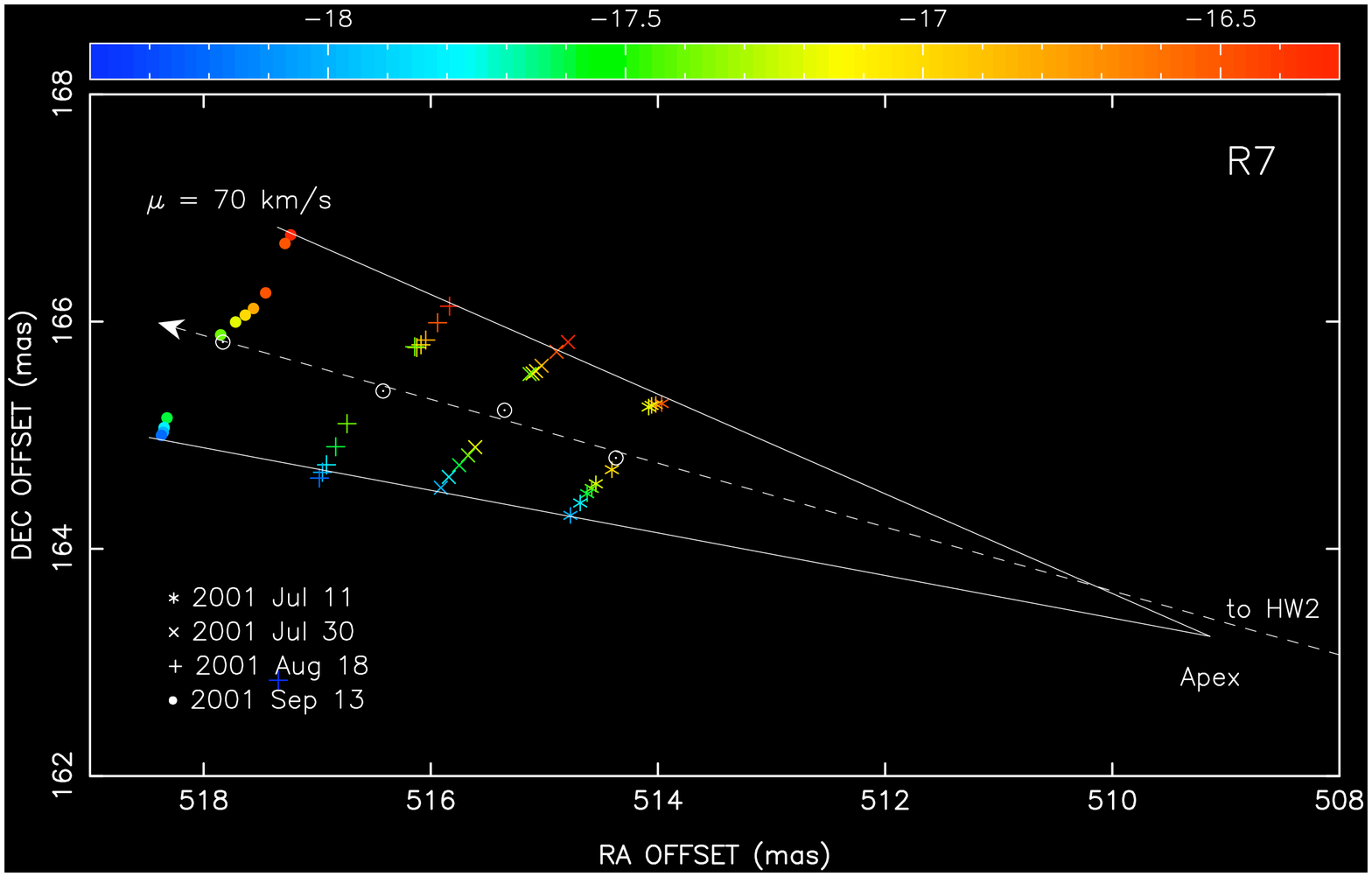} 
 \caption{Fan-like water maser structure  formed by the expanding motions of the R7 linear structure at scales of $\sim$ 4 mas (3 AU). The continuum lines indicate the linear fitting of the outer water masers of the fan-like structure (see \S~4). The dashed line indicates the linear fitting of the average positions (open circles) of each of the linear structures observed in the different epochs. This central line points to HW2 within $\sim$ 20 mas. The apex of the fan-like structure is almost exactly on this line (within $\sim$ 0.3 mas), but much closer to the position of the detected maser emission than to that of HW2. The water maser structure, while moving toward the north-east, presents a LSR radial velocity gradient (colour code in km~s$^{-1}$) along the linear structure, suggesting that it is caused by a rotation (see \S~4).}
\end{figure}

The VLBA multi-epoch water maser observations towards the HW2-disk system show now, for the
first time in a high-mass  YSO, and similar to what is
observed in low-mass YSOs, the simultaneous presence of a high
velocity jet surrounded by a wide-angle outflow in the HW2-disk system
at scales $\la$ 600~AU.  This is inferred from the observation
of the R6, R7, and R8 masers (\S 3.1). 
These masers are located (in projection) near the north-east edge of the HW2 dust disk, as observed
with the SMA by Patel et al. (2005). Their location and motions suggest that they are
associated with the blue-shifted, north-east lobe of the HW2 bipolar molecular (HCO$^+$) outflow (G\'omez et al. 1999), tracing
wide-angle movements. In fact, while R6
and R8 (which are located at both sides of R7; Figures 1 and 2)
exhibit motions in the sky almost in opposite directions (R6 toward
the south-east with PA $\simeq$ 118$^{\circ}$, and  R8 toward the
north-west with PA $\simeq$ --35$^{\circ}$), R7 shows faster motions
toward the north-east (PA $\simeq$ 74$^{\circ}$), along  a direction closer to the jet axis, 
with a difference in position angle of $\sim$ 30$^{\circ}$ with respect to the highly collimated HW2 radio jet (PA $\simeq$ 45$^{\circ}$) (see Figures 1 and 2). All
these properties can be explained in a shock interaction scenario
within the disk-HW2-outflow system. In this scenario, outlined in
Figure 5, we propose that R6 and R8 represent emission fronts
from the walls of expanding cavities of the circumstellar gas around HW2,  created and shocked
by a wide-angle primary wind from this source (outflow opening angle
$\sim$ 102$^{\circ}$). We also propose that the linear structure R6 (V$_{LSR}$ $\simeq$ --11 to --18~km~s$^{-1}$, blue-shifted with respect to the ambient cloud; \S 3.1)
is the counterpart of the R1-R3 linear structure (V$_{LSR}$ $\simeq$ --10 to +1~km~s$^{-1}$, red-shifted; T2001b) located in the opposite side
(to the west of HW2; Figure 1). This structure was interpreted previously as produced
in the shocked walls of an expanding cavity (T2001b). In fact,   although both linear
structures (R6 and R1-3) have similar orientations,  their proper
motions are almost in opposite directions (R6: $\mu$ $\simeq$ 13~km~s$^{-1}$
at PA $\simeq$ 118$^{\circ}$; R1: $\mu$ $\simeq$ 5 km~s$^{-1}$ at
PA $\simeq$ --55$^{\circ}$). Furthermore, in R6 there is a LSR radial velocity gradient of $\sim$ 4-5 km~s$^{-1}$ along the filament,
with more blue-shifted velocities away from HW2 (Fig. 2), and in R1 there
is also a LSR radial velocity gradient of $\sim$ 2~km~s$^{-1}$ along the
filament with the same trend, more red-shifted velocities away from HW2 (T2001b). These velocity gradients with the same trend in R6 and R1 can be explained by the acceleration of maser gas at the walls of the cavities by the wide-angle primary wind from HW2.

On the other hand, the higher
motions of the R7 masers along  a direction closer to the main axis of the HW2 radio jet suggest that the interaction
of the wide-angle primary wind from HW2 with the molecular gas is within the cavities,
through a less slowed-down ambient medium.
As seen in \S 3.1 and Figure 6, the expanding motions of the masers form a fan-like
structure at very small  scales ($\sim$ 4 mas, $\sim$ 3
AU).  By means of a linear fitting of the average positions of each of the
linear structures observed in the different epochs of R7, we find that
the line joining these average positions (hereafter central line; Fig. 6) is pointing back toward the HW2 position within $\sim$
20 mas. A linear fitting of the positions of each of the two edges of the R7 structures at different epochs gives two lines that intersect
at ($\Delta$$\alpha$,$\Delta$$\delta$)
$\simeq$ (509.1 mas,163.2 mas),  almost exactly (within $\sim$ 0.3 mas!) on the central line, but much closer to the position of the detected maser emission than  to that of HW2. From these values we estimate a dynamical time-scale
of $\sim$ 0.3 yr for the expanding structure. These results suggest that the R7 masers, which are moving along a direction
with a difference angle of $\sim$ 30$^{\circ}$ with respect to the main axis of the radio jet, are excited by the HW2 wide-angle primary wind via interaction with small clumps
of gas within the cavity, accelerated and beginning to expand in a
short-lived event. In fact, we note that R7 is not detected
in the fifth epoch of the VLBA 2001-2002 data set, six months after
its first detection, consistent with the short dynamical time.

\begin{figure}
 \includegraphics[scale=.7]{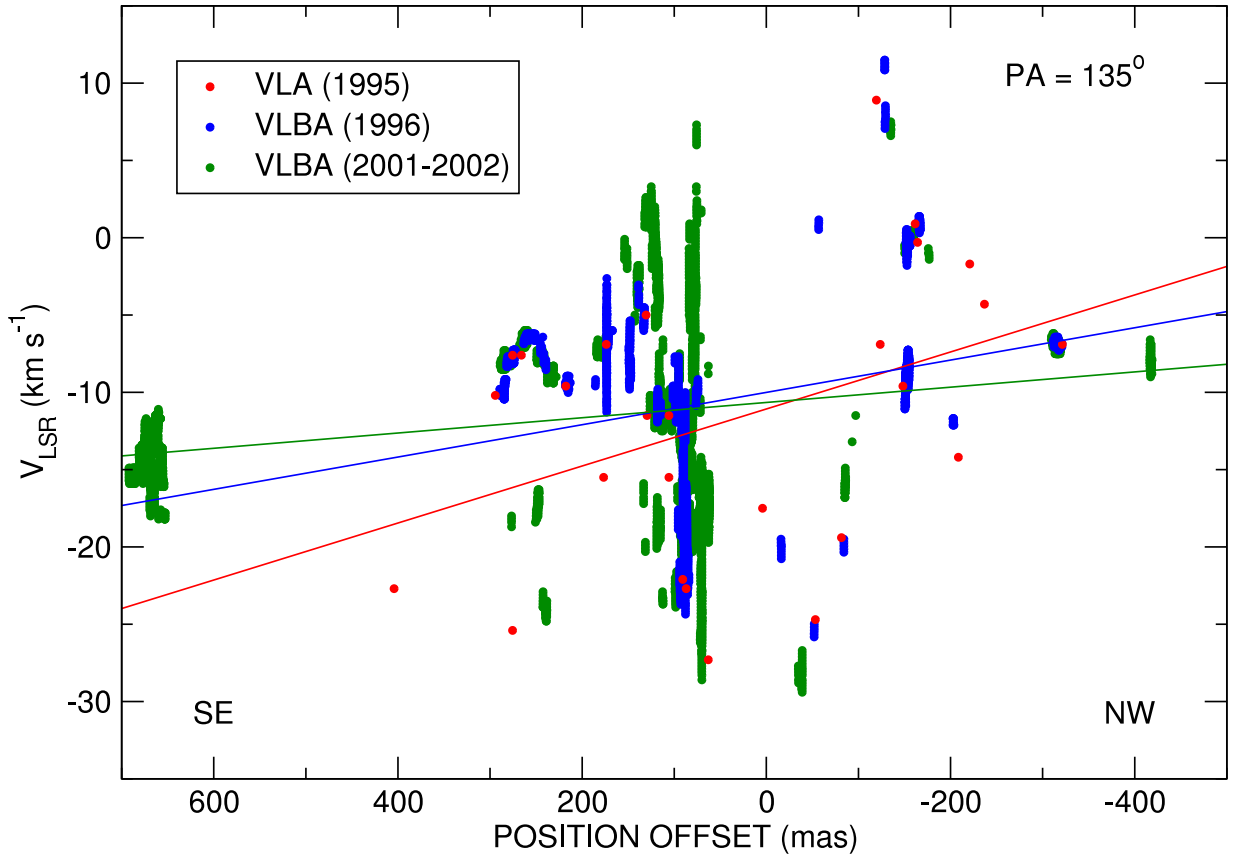} 
 \caption{Position-LSR radial velocity distribution of the H$_2$O masers as observed with the VLA 1995 (red dots), VLBA 1996 (blue dots), and VLBA 2001-2002 (green dots) data sets along an axis with PA = 135$^{\circ}$ (perpendicular to the main axis of the HW2 radio jet). Position offsets are with respect to the HW2 position. Masers with a broad velocity distribution ($\sim$ 30~km~s$^{-1}$) with position offsets $\sim$ 70 mas
 correspond to the R4 structure. Masers associated with R5 ($\sim$ 0.6$''$ south of HW2) have not been plotted here. Colour lines indicate the linear fitting of the position-velocity distribution of all the masers for the VLA 1995 (red), VLBA 1996 (blue), and VLBA 2001-2002 (green) data sets. A velocity gradient is observed in the VLA 1995 data set, with more red-shifted velocities toward the north-west. This gradient, although still present, is less evident in the VLBA 1996 and VLBA 2001-2002 data sets  (see \S 4).}
\end{figure}

In addition, the LSR radial velocity gradient
of $\sim$ 2 km s$^{-1}$ observed along the linear structure
of R7, with more red-shifted velocities towards the north-western part
(Figures 2 and 6), could trace a rotating outflow from HW2, extracting
angular momentum from the disk. In fact, a LSR radial velocity gradient  of $\sim$
5~km~s$^{-1}$ is observed along the major axis of the rotating disk
of HW2, with more red-shifted velocities also toward the north-west
(Patel et al. 2005, Jim\'enez-Serra et al. 2007, 2009). This LSR radial velocity gradient was also observed in the H$_2$O masers of the VLA 1995 data set (T96), with more red-shifted velocities toward the north-west, although it is less evident
in the VLBA 1996 and VLBA 2001-2002 data sets (Fig. 7).
However, with the present data we
cannot distinguish whether the observed velocity gradient  in the linear
structures of R7 is due to a rotating outflow  or due to
intrinsic velocity shifts in the  surrounding ambient molecular gas, before
the interaction with the  HW2 primary wind. To differentiate clearly between
these two possibilities it would be necessary  to detect another
fan-like water maser structure in the opposite side, to the
south-west of HW2, with similar velocity gradients. We also note that even if the velocity gradient observed in the masers of R7 has a rotation 
origin, 
it  seems that now the structure is not conserving angular momentum, because even when it expands 
significantly between the four epochs, the total velocity difference for any
given epoch remains approximately constant (see Fig. 6). If we had a structure with angular
momentum conservation this total velocity difference would have to diminish with
expansion.  Most likely, the features constituting R7 are at present moving ballistically.

The scenario proposed to explain the R6, R7, R8, and R1-3 structures in HW2 is similar to what
was proposed by Moscadelli, Cesaroni, \& Rioja (2005) to explain the conical outflow in the high-mass object IRAS 20126+4104 (also observed through VLBI multi-epoch water maser observations). The main difference is that while in IRAS 20126+4104 the masers were observed to be tracing an outflow with an opening angle of $\sim$ 34$^{\circ}$, in HW2 the masers trace a wider outflow (opening angle of $\sim$ 102$^{\circ}$) with expanding cavities surrounding the radio continuum jet (opening angle of $\sim$ 18$^{\circ}$).

We also considered the possibility that the R6, R8, and R1-3 masing regions
could be associated with the far wings of a
bow shock driven by a collimated jet (i.e., the
jet observed in radio continuum) with successive working surfaces
produced by an outflow variability, in the absence of a wide-angle wind, as proposed for the low mass protostar XZ Tau (Krist et al. 2008). A working surface is formed in a supersonic flow
(in this case, the jet) as a result of a supersonic increase in its ejection velocity. High
velocity material catches up with the previously ejected, lower
velocity flow, forming two shocks: a leading one which accelerates the
lower velocity flow and a reverse shock which decelerates the higher
velocity material ejected at later times (for details see Raga et al. 1990).
Material is constantly being accumulated between the two shocks, and
part of it is ejected sideways, transferring its linear momentum
to the surrounding medium (by means of a bow shock driven into
this medium). However, although we cannot rule out this possibility
(detailed modelling of the proper motions are required),
the motions of the R7 structure
along an axis at an angle
of $\sim$ 30$^{\circ}$ with respect to the radio jet axis would imply the presence of a wider outflow
than that indicated by the jet.

Wide-angle outflows have been observed in YSOs of all luminosities, from $\sim$ 1 to 10$^5$ L$_{\odot}$ (e.g., Shepherd 2005; Qiu et al. 2009; Matthews et al. 2010). 
What makes our VLBA water maser results in Cepheus A HW2 highly significant is 
the fact that both, highly collimated and wide-angle outflows, are observed 
simultaneously, and at similar physical scales. Although at the 
moment, there are no 
models predicting the simultaneous presence of these two  kinds of
outflows in young high-mass stars, as those developed for low-mass
YSOs, we think that our observational results represent an important constraint 
for future models on high-mass stars when trying to reproduce
the scale where these outflows are observed. Future (sub)mm high-angular observations would also be very valuable to look for the presence of molecular (e.g., CO, SiO) conical shells associated with HW2 to have an even stronger comparison to what is observed in low-mass objects.

\section{Conclusions}

We have carried out VLBA multi-epoch (2001-2002 epochs) water maser observations toward the high-mass protostar Cepheus A HW2 with 0.4 mas (0.3 AU) resolution. These observations show that the previously 
reported R5 and R4 arcuate structures (1996 epochs; T2001b) have suffered large morphological changes at scales of $\la$ 0.1$''$ (70 AU) in a time span of five years. 
In particular, the R5 expanding bubble structure observed five years before, located  $\sim$ 0.6$''$ (400 AU) south of HW2, is currently dissipating in the circumstellar medium, losing its previous degree of symmetry,  thus corroborating the very short-lived nature of this phenomenon. On the other hand, the masers of the R4 structure trace a nearly elliptical patchy expanding ring of $\sim$ 70 mas size (50 AU), supporting the ring scenario proposed previously by Gallimore et al. (2003), rather than the bow-shock scenario proposed by T2001b.  We propose that the central YSO  driving
the observed expanding motions of the ring is a still unidentified source  (probably a massive object given the high luminosity of the associated water masers) that should be located $\sim$ 0.2$''$ (145 AU) south of HW2. 

Our water maser observations also reveal a relatively slow wide-angle outflow (opening angle $\sim$ 102$^{\circ}$) with the simultaneous
presence of the high-velocity ionised jet (opening angle $\sim$ 18$^{\circ}$)  associated with HW2. The presence of high-velocity jets, enclosed within wide-angle outflows, has been observed in low-mass protostars. The importance of our result is that 
this is now observed in a massive protostar, with the highly collimated and wide-angle outflow in HW2 occurring 
simultaneously at a similar physical scale of $\sim$ 1$''$ (700 AU). Our results provide important constraints for future models of high-mass stars that try to reproduce the different outflow opening angles observed in the Cepheus A HW2 system.

The high brightness of the masers  has proven to be extremely useful to study the gas very 
close to young massive stars through VLBI multi-epoch observations. These observations, in addition, are
revealing very short-lived events. Particularly, in Cepheus A HW2, the different water maser structures detected around this source have dynamical time-scales from 0.3 to 30 yr, with large structural changes in the time span of years at scales of $\la$ 0.1$''$ (70 AU), providing 
new insights in the study of the dynamic scenario that seems to characterise
the formation of high-mass stars.

\section*{Acknowledgments}

We would like to thank our referee for the very careful and useful report on our manuscript. GA, RE, JFG, and JMT acknowledge support from MICINN (Spain) AYA2008-06189-C03 grant (co-funded with FEDER funds). GA, JFG, and JMT acknowledge support from Junta de Andaluc\'{\i}a (Spain). GG acknowledges support from projects FONDAP No. 15010003 and BASAL PFB-06. SC acknowledges support from CONACyT grant 60581.
JC and AR acknowledge support from CONACyT grant 61547. LFR acknowledges the support
of DGAPA, UNAM, and of CONACyT (M\'exico).
WV acknowledges support by the Deutsche Forschungsgemeinschaft (DFG) through the Emmny Noether Research grant VL 61/3-1.

\section*{Supporting information}

We provide in the form of electronic animations the water maser motions observed in R6 (Anim-R6.gif, Anim-R6-i.gif, Anim-R6-ii.gif), R7 (Anim-R7.gif), R8 (Anim-R8.gif), R1 (Anim-R1.gif, Anim-R1-i.gif, Anim-R1-ii.gif),  R5 (Anim-R5.gif, Anim-R5-ab.gif, Anim-R5-cde.gif, Anim-R5-f.gif) and R4 (Anim-R4.gif, Anim-R4-zoom.gif).
Colour code indicates the LSR radial velocity (km~s$^{-1}$) of the maser spots.

\bsp

\label{lastpage}

\end{document}